\RequirePackage{lineno}
\documentclass[preprint,amsmath,amssymb,superscriptaddress,unsortedaddress]{revtex4}

\usepackage{epsf}
\usepackage[dvipdfmx]{graphicx}
\usepackage{color}
\usepackage[dvipdfmx]{hyperref}
\usepackage{newfloat}

\begin{document}

\title{Spintronic superconductor in a bulk layered material \\ with natural spin-valve structure}

\author{Shunsuke Sakuragi\text{*}}
\affiliation{Institute for Solid State Physics, University of Tokyo, Kashiwa, Chiba, Japan}

\author{S.~Sasaki} 
\affiliation{Department of Physics, The University of Tokyo, Tokyo, Japan}

\author{R.~Akashi}
\affiliation{Department of Physics, The University of Tokyo, Tokyo, Japan}

\author{R.~Sakagami} 
\affiliation{Department of Applied Physics and Phisico-Informatics, Keio University, Yokohama, Japan}

\author{K.~Kuroda} 
\affiliation{Institute for Solid State Physics, University of Tokyo, Kashiwa, Chiba, Japan}

\author{C.~Bareille}
\affiliation{Institute for Solid State Physics, University of Tokyo, Kashiwa, Chiba, Japan}

\author{T.~Hashimoto}
\affiliation{Institute for Solid State Physics, University of Tokyo, Kashiwa, Chiba, Japan}

\author{T.~Nagashima}
\affiliation{Institute for Solid State Physics, University of Tokyo, Kashiwa, Chiba, Japan}

\author{Y.~Kinoshita}
\affiliation{Institute for Solid State Physics, University of Tokyo, Kashiwa, Chiba, Japan}

\author{Y.~Hirata} 
\affiliation{Institute for Solid State Physics, University of Tokyo, Kashiwa, Chiba, Japan}
\affiliation{Department of Applied Physics, National Defense Academy of Japan, Yokosuka, Japan}

\author{M.~Shimozawa} 
\affiliation{Institute for Solid State Physics, University of Tokyo, Kashiwa, Chiba, Japan}
\affiliation{Division of Materials Physics, Graduate School of Engineering Science, Osaka University, Toyonaka, Japan}

\author{S.~Asai} 
\affiliation{Institute for Solid State Physics, University of Tokyo, Kashiwa, Chiba, Japan}

\author{T.~Yajima} 
\affiliation{Institute for Solid State Physics, University of Tokyo, Kashiwa, Chiba, Japan}

\author{S.~Doi} 
\affiliation{Department of Physics, The University of Tokyo, Tokyo, Japan}
\affiliation{Research Center for Computational Design of Advanced Functional Materials, National Institute of Advanced Industrial Science and Technology, Tsukuba, Japan.}

\author{N.~Tsujimoto} 
\affiliation{Department of Physics, The University of Tokyo, Tokyo, Japan}

\author{S.~Kunisada} 
\affiliation{Institute for Solid State Physics, University of Tokyo, Kashiwa, Chiba, Japan}

\author{R.~Noguchi} 
\affiliation{Institute for Solid State Physics, University of Tokyo, Kashiwa, Chiba, Japan}

\author{K.~Kurokawa} 
\affiliation{Institute for Solid State Physics, University of Tokyo, Kashiwa, Chiba, Japan}

\author{N.~Azuma} 
\affiliation{Department of Applied Physics and Phisico-Informatics, Keio University, Yokohama, Japan}

\author{K.~Hirata} 
\affiliation{Department of Applied Physics and Phisico-Informatics, Keio University, Yokohama, Japan}

\author{Y.~Yamasaki} 
\affiliation{Research and Services Division of Materials Data and Integrated System, National Institute for Materials Science, Tsukuba, Japan}
\affiliation{RIKEN Center for Emergent Matter Science, Wako, Japan}
\affiliation{PRESTO, Japan Science and Technology Agency, Kawaguchi, Japan}

\author{H.~Nakao} 
\affiliation{Condensed Matter Research Center and Photon Factory, Institute of Materials Structure Science, High Energy Accelerator Research Organization, Tsukuba, Japan}

\author{T.~K.~Kim} 
\affiliation{Diamond Light Source, Harwell Campus, Didcot, UK}

\author{C.~Cacho} 
\affiliation{Diamond Light Source, Harwell Campus, Didcot, UK}

\author{T.~Masuda} 
\affiliation{Institute for Solid State Physics, University of Tokyo, Kashiwa, Chiba, Japan}

\author{M.~Tokunaga} 
\affiliation{Institute for Solid State Physics, University of Tokyo, Kashiwa, Chiba, Japan}

\author{H.~Wadati} 
\affiliation{Institute for Solid State Physics, University of Tokyo, Kashiwa, Chiba, Japan}
\affiliation{Graduate School of Material Science, University of Hyogo, Hyogo, Japan}

\author{K.~Okazaki} 
\affiliation{Institute for Solid State Physics, University of Tokyo, Kashiwa, Chiba, Japan}

\author{S.~Shin} 
\affiliation{Institute for Solid State Physics, University of Tokyo, Kashiwa, Chiba, Japan}
\affiliation{TheUniversity of Tokyo, Kashiwa, Chiba, 277-8581, Japan}

\author{Y.~Kamihara} 
\affiliation{Department of Applied Physics and Phisico-Informatics, Keio University, Yokohama, Japan}  
\affiliation{Center for Spintronics Research Network, Keio University, Yokohama, Japan}

\author{Minoru~Yamashita\text{*}} 
\affiliation{Institute for Solid State Physics, University of Tokyo, Kashiwa, Chiba, Japan}

\author{Takeshi~Kondo\text{*}}
\affiliation{Institute for Solid State Physics, University of Tokyo, Kashiwa, Chiba, Japan}

\date{\today}

\maketitle
{\bf Multi-layered materials provide fascinating platforms to realize various functional properties,  possibly leading to future electronic devices controlled by external fields. In particular, layered magnets coupled with conducting layers have been extensively studied recently for possible control of their transport properties via the spin structure. Successful control of quantum-transport properties in the materials with antiferromagnetic (AFM) layers, so-called natural spin-valve structure, has been reported for  the Dirac Fermion and  topological/axion materials \cite{MasudaSciAdv,BorisenkoNComm, ZhangPRL, RienksNature, LiCondmat, LiSciAdv, huCondmat, MaSciAdv, YeNature, YinNature, KangNMat}. However, a bulk crystal in which magnetic and superconducting layers are alternately stacked has not been realized until now, and the search for functional properties in it is an interesting yet unexplored field in material science. Here, we discover superconductivity providing such an ideal platform in EuSn$_2$As$_2$ with the van der Waals stacking of magnetic Eu layers and superconducting Sn-As layers, and present the first demonstration of a natural spin-valve effect on the superconducting current. Below the superconducting transition temperature ($T_c$), the electrical resistivity becomes zero in the in-plane direction. In contrast, it, surprisingly, remains finite down to the lowest temperature in the out-of-plane direction, mostly due to the structure of intrinsic magnetic Josephson junctions in EuSn$_2$As$_2$. The magnetic order of the Eu layers (or natural spin-valve) is observed to be extremely soft, 
allowing one to easy control of the out-of-plane to in-plane resistivities ratio from 1 to $\infty$
by weak external magnetic fields. The concept of multi-functional materials with stacked magnetic-superconducting layers will open a new pathway to develop novel spintronic devices with magnetically controllable superconductivity.}

EuSn$_2$As$_2$ has a structure consisting of the magnetic insulating Eu layers and the conducting Sn-As layers, which are mutually coupled via the van der Waals force (Fig. 1a) \cite{Arguilla, Sakagami}. The magnetic interactions in the intra- and inter-layers of Eu ions have been suggested to be ferromagnetic and antiferromagnetic, respectively \cite{Arguilla}.  The X-ray magnetic scattering (Fig. 1b) confirms that the antiferromagnetic order of Eu moment develops along the $c$-axis below the N\'eel temperature ($T_N$) of $\sim 23$~K. Intriguingly, this compound exhibits a soft ferromagnetic behavior in the MH curves (Fig. 1c) with a very strong magnetic response, unlike a typical antiferromagnet. The ferromagnetic direction, thus, can be changed by small external magnetic fields, enabling one to easily control the spin-valve effect of the Eu layers. We have observed the formation of magnetic domains with the size of $\sim 50$~$\mu$m in the $a$-$b$ plane below $T_N $ (the left panel of Fig. 1d) by the polarizing microscopy, which visualizes the magnetic domain. This image was taken in zero magnetic fields, where the Eu spins should be antiferromagnetic with no net magnetization on average over the whole bulk crystal. Hence, the observed signal of magnetic domain should come from the optical anisotropy of antiferromagnetic domains with canted order inherent in the Eu layer \cite{LiCondmat}. 

Most importantly, we find that EuSn$_2$As$_2$ shows superconductivity, thus it is the first material with van der Waals stacked magnetic-superconducting layers. As a fascinating property for it, we present the natural spin-valve effect both on the normal and superconducting electrons, which is observed in the temperature dependence of resistivities (Fig. 2a) along the $ab$-  and $c$-axis ($\rho_{ab}$ and $\rho_{c}$, respectively). 

While both the resistivities show metallic behaviors above $T_N$ (Fig. 2a), $\rho_{c}$ is significantly increased below $T_N$, demonstrating a two-dimensional (2D) confinement of the conduction electrons by the spin-valve effect \cite{MasudaSciAdv}. We found the superconducting transition in $\rho_{ab}$ at the temperature ($T_c$) of 4.8~K. The  $\rho_{ab}(T)$ has a two-step feature on cooling below $T_c$ with a shoulder-like structure (at $T_1$) before becoming zero below $T_0 \sim 0.8$~K (see the inset of Fig. 2a and the zero-field data of Fig. 2c). The two-step transition has been confirmed for several samples from different crystal-batches, thus it should be an intrinsic feature of EuSn$_2$As$_2$; notably, the similar transition seems to be often observed in 2D superconductors \cite{EleyNPhys,ChenNCom}. 

The bulk superconductivity is also confirmed by the Meissner effect (Fig. 2d); diamagnetism due to the superconductivity is observed as a sharp decrease of the magnetization below $T_c$, and it eventually disappears with increasing magnetic field. Note here that the overall shift for each curve in Fig. 2d comes from the magnetic susceptibility of Eu layers. 
The critical magnetic field ($H_{c2}$) is found to be very small ($\sim 40$~mT) and isotropic (Fig. 2c and e). While these magnetic properties are usually obtained for the conventional type-I superconductors, our resistivity data exhibit a clear anomaly against it: $\rho_{c}$ becomes abruptly enhanced below $T_c$ (Fig. 2a), giving rise to an infinite ratio of resistivities ($\rho_{c}$/$\rho_{ab}$) at low temperatures. The increase of $\rho_{c}$ below $T_c$ is found to be suppressed by a weak magnetic field close to $H_{c2}$ (Fig. 2b), indicating that the additionally enhanced resistance along the $c$-axis is caused by the bulk superconductivity. The extremely anisotropic resistance is apparently unfeasible in the type-I superconductivity, thus highly unconventional superconductivity should be realized in EuSn$_2$As$_2$.  
Since it is also in sharp contrast to the typical type-II superconductivity observed in the related nonmagnetic material NaSn$_2$As$_2$ \cite{GotoJPSJ, IshiharaPRB}, the magnetic Eu layers are implied to play an essential role in the anisotropic superconductivity and the superconducting spin-valve effect of this compound, capable of tuning the $\rho_{c}$/$\rho_{ab}$ ratio between $\infty$ to 1 by weak external magnetic fields.

In order to clarify the mechanism of the abnormal resistivities, we have investigated the electronic structure of EuSn$_2$As$_2$ by the synchrotron-based angle-resolved photoemission spectroscopy (ARPES) together with the DFT calculations. The Femi surface mapping over multiple Brillouin zones across the $\Gamma$ point (Fig. 3b) reveals that the Fermi surfaces are formed only around $k_x=k_y=0$. The overall ARPES dispersion (Fig. 3c) along the $\Gamma$-M cut (a red line in Fig.  3b) is well reproduced by our DFT calculations slightly hole-doped (blue curves in Fig. 3c). 

The calculations predict that a Lifshitz transition occurs across $T_N$ due to the hybridization between the original bands and those folded about the antiferromagnetic zone boundaries, which reconstruct three-dimensional (3D) Fermi surfaces with a peanuts-shape above $T_N$ (left panel in Fig. 3a) into two types of Fermi surfaces with 2D and 3D cylindrical shapes below $T_N$ (noted by $\alpha$ and $\beta$ in the right panel of Fig. 3a, respectively). Our ARPES data (Fig. 3d and e) indeed agree with this prediction. The top panels of Fig. 3d plot the Fermi surface mapping along a $k_z-k_y$ sheet obtained at $k_x=0$ by widely changing photon energies (or the $k_z$ value); the left and right are the data obtained above and below $T_N$, respectively. In the bottom panels, the mapping along the $k_x-k_y$ sheet across $\Gamma$ is extracted. The largest Fermi surface (noted as $\gamma$ in Fig. 3d), absent in our calculations, should be for surface states \cite{LiCondmat, GuiACS}, since it is 2D-like showing no dispersion along $k_z$ regardless of measured temperature; the spectral intensities modulate with photon energies due to the matrix element effect in photoemission. The drastic variation across $T_N$ for bulk states is seen close to the momentum line at $k_x=k_y=0$, being consistent with our DFT calculations (yellow dashed lines in Fig. 3d). This is more clearly demonstrated in Fig. 3e, by extracting the momentum distribution curves from the $k_z$ dispersion of Fig. 3d: a 3D-like dispersion above $T_N$ is changed to be 2D-like below $T_N$ with almost no variation in $k_y$ position (marked by black bars) along $k_z$ (Fig. S 4). Importantly, this Lifshitz transition should be the cause of the increase in $\rho_{c}$ below $T_N$.

More detailed properties of the band structure can be obtained from the DFT-based analysis. The total spin states should degenerate in the AFM phase of EuSn$_2$As$_2$. 
However, we found that the top valence-band formed by the electrons spatially adjacent to the Eu layer become spin polarized concurrently with the band split in the AFM phase (Fig. 4b, c, Fig. S4b) if we see the spin component within only half a unit cell, which is equivalent to the projection of electronic state to the direction of the adjacent Eu layer. This is due to the proximity of exchange interaction, which lifts the spin degeneracy, and thus the up- and down-spin states are flipped in energy for the electrons nearby the Eu layer with the opposite ferromagnetism (Fig. 4a). Consequently, the spin-polarized electrons are confined within the block sandwiched by the Eu layers with the same spin; a single Eu layer acts as a spin filter, and thus the stacking of the layer in a crystal forms a natural spin-valve. 
Our ARPES data, therefore, exhibit the natural spin-valve effect on the band structures.

To investigate the superconducting state of EuSn$_2$As$_2$, we have used ARPES with a laser ($h\nu$=7 eV), capable of measurements with high energy and momentum resolutions. Figure 3f shows the band dispersion measured by the laser-based ARPES above $T_c$ ($T$=6 K); the measured momentum cut is represented in Fig. 3b (a blue line), while the measured $k_z$ is not zero at $h\nu$=7 eV. We observe three bands, assigned as the band of surface states ($\gamma$-band), and two other bulk bands derived from the Sn-As orbitals: a two-dimensional M-shaped band ($\alpha$-band) and a three-dimensional hole-like band ($\beta$-band) (Fig. S5). 

In Fig. 3g, we examine the temperature evolution of the energy distribution curves (EDCs) at $k_F$ for the hole-like band. The energy shift of the leading-edge midpoint is clearly seen in the spectra below $T_c$, and the gap magnitude increases on cooling down to the lowest temperature. The experiments were repeated for different pieces of samples, and similar results were always obtained. In Fig. 3h, we estimate the superconducting gap by fitting the EDCs to the Bardeen-Cooper-Schrieffer (BCS) spectral function \cite{OtaPRL}. In the same panel, we also overlap the superconducting gap (a dotted red curve reaching $\Delta$ $\sim$ 0.5 meV at 0 K) obtained from the BCS theory for a superconductor with $T_c$ = 4.8 K, same as that of EuSn$_2$As$_2$. A good agreement between the two is obtained (Fig. S 6).

The anomalous phenomena we found for EuSn$_2$As$_2$ are mainly the following three: (1) small $H_{c2}$, (2) two-step superconducting transition in $\rho_{ab}$, and (3) non-zero $\rho_{c}$ below $T_c$. 
The small $H_{c2}$ [the finding (1)] could be explained by the magnetic proximity effect of the Eu layer on the superconducting As-Sn layer. In our calculations, the valence band of each Sn-As sheet splits the energies of up- and down-spin by $\sim$ 0.1 eV due to this effect (Fig. 4b and c; Fig. S 4b), which corresponds to the internal magnetic field of $\sim$ 2000 T \cite{SaitoNPhys}. Since the ferromagnetic Eu-spin is observed to be very soft and isotropic, small external magnetic fields in any direction can generate a large internal magnetic field, which substantially reduces $H_{c2}$. 
Moreover, the Cooper pairs formed in the spin-split state (see Fig. 4a) could be quite sensitive to the local magnetic environment, which may give rise to a small Pauli-limiting $H_{c2}$; this scenario is consistent with the small and isotropic $H_{c2}$ observed in EuSn$_2$As$_2$.

The two-step superconducting transition [the finding (2)] has been previously observed in various 2D superconductors \cite{EleyNPhys,ChenNCom}. A similar situation may occur in the in-plane resistivity ($\rho_{ab}$) of EuSn$_2$As$_2$: Bellow $T_c$, superconducting puddles emerge only locally in the As-Sn-Sn-As blocks, leaving a finite resistance in the in-plane electrical current ($\sim T_1$ in Fig. 2a, Fig. 4d). On further cooling, the phase coherence eventually percolates into the whole in-plane, turning  $\rho_{ab}$ to be zero ($\sim T_0$ in Fig. 2a, Fig. 4e). 

The non-zero $\rho_{c}$ below $T_c$ [the finding (3)] is more puzzling. This, however, could be understood by regarding the current system as a 3D array of Josephson junctions with inhomogeneous magnetic domains (Fig. 4f). 
A junction made of the magnetic insulator can shift the superconducting phase by $\pi$ across it, even if it consists only of a single atomic layer \cite{KawabataPRL}. The phase shifts either of 0 or $\pi$ is determined by the width of the magnetic layer, the energy of exchange splitting, and other subtle features of the band structure \cite{KawabataLTP}. In EuSn$_2$As$_2$, magnetic domains working as a phase shifter are changed for different paths of superconducting electrons. This could cancel the total $I_c$ along the $c$-axis, as in the device of DC-superconducting quantum interference (Fig. 4g), or possibly scramble the interlayer phase coherence. Consequently, only the current of normal electrons ($I_n$) excited by the thermal fluctuation and impurities contribute to the total current, generating a finite resistance, which could even increase below $T_c$. Nevertheless, more detailed analyses would be required as future researches for the full description of the fascinating phenomena we found in EuSn$_2$As$_2$.

Various applications are conceivable for the physical phenomena observed in EuSn$_2$As$_2$ with intrinsic magnetic Josephson junctions. Fascinatingly, the value of $\rho_{c}$/$\rho_{ab}$ can be tuned between $\infty$ to 1 by a small magnetic field, owing to the exceptionally soft magnetism of the Eu layers. In addition, we could select the magnetic junction to shift the superconductive phase between 0 and $\pi$ by controlling the magnetic domains of the Eu layers, such as by detwinning with a piezo substrate. This will open the way to develop new types of magnetic memory showing the signal of 0 (0-0 junction) or $\infty$ (0-$\pi$ junction, shown in Fig. 4f), enabling an ultrafast processing specific of antiferromagnetism \cite{BaltzRMP,NemecNPhys}. Most importantly, our results provide a novel concept of multi-functional materials with magnetic-superconducting layers alternately stacked, leading to a new avenue of applications with van der Waals materials. 

\clearpage

\bibliography{ref}

\begin{thebibliography}{32}
\expandafter\ifx\csname natexlab\endcsname\relax\def\natexlab#1{#1}\fi
\expandafter\ifx\csname bibnamefont\endcsname\relax
  \def\bibnamefont#1{#1}\fi
\expandafter\ifx\csname bibfnamefont\endcsname\relax
  \def\bibfnamefont#1{#1}\fi
\expandafter\ifx\csname citenamefont\endcsname\relax
  \def\citenamefont#1{#1}\fi
\expandafter\ifx\csname url\endcsname\relax
  \def\url#1{\texttt{#1}}\fi
\expandafter\ifx\csname urlprefix\endcsname\relax\def\urlprefix{URL }\fi
\providecommand{\bibinfo}[2]{#2}
\providecommand{\eprint}[2][]{\url{#2}}

\bibitem[{\citenamefont{Masuda et~al.}(2016)\citenamefont{Masuda, Sakai,
  Tokunaga, Yamasaki, Miyake, Shiogai, Nakamura, Awaji, Tsukazaki, Nakao
  et~al.}}]{MasudaSciAdv}
\bibinfo{author}{\bibfnamefont{H.}~\bibnamefont{Masuda}},
  \bibinfo{author}{\bibfnamefont{H.}~\bibnamefont{Sakai}},
  \bibinfo{author}{\bibfnamefont{M.}~\bibnamefont{Tokunaga}},
  \bibinfo{author}{\bibfnamefont{Y.}~\bibnamefont{Yamasaki}},
  \bibinfo{author}{\bibfnamefont{A.}~\bibnamefont{Miyake}},
  \bibinfo{author}{\bibfnamefont{J.}~\bibnamefont{Shiogai}},
  \bibinfo{author}{\bibfnamefont{S.}~\bibnamefont{Nakamura}},
  \bibinfo{author}{\bibfnamefont{S.}~\bibnamefont{Awaji}},
  \bibinfo{author}{\bibfnamefont{A.}~\bibnamefont{Tsukazaki}},
  \bibinfo{author}{\bibfnamefont{H.}~\bibnamefont{Nakao}},
  \bibnamefont{et~al.}, \bibinfo{journal}{Science Advances}
  \textbf{\bibinfo{volume}{2}} (\bibinfo{year}{2016}),
  \eprint{https://advances.sciencemag.org/content/2/1/e1501117.full.pdf},
  \urlprefix\url{https://advances.sciencemag.org/content/2/1/e1501117}.

\bibitem[{\citenamefont{Borisenko et~al.}(2019)\citenamefont{Borisenko,
  Evtushinsky, Gibson, Yaresko, Koepernik, Kim, Ali, van~den Brink, Hoesch,
  Fedorov et~al.}}]{BorisenkoNComm}
\bibinfo{author}{\bibfnamefont{S.}~\bibnamefont{Borisenko}},
  \bibinfo{author}{\bibfnamefont{D.}~\bibnamefont{Evtushinsky}},
  \bibinfo{author}{\bibfnamefont{Q.}~\bibnamefont{Gibson}},
  \bibinfo{author}{\bibfnamefont{A.}~\bibnamefont{Yaresko}},
  \bibinfo{author}{\bibfnamefont{K.}~\bibnamefont{Koepernik}},
  \bibinfo{author}{\bibfnamefont{T.}~\bibnamefont{Kim}},
  \bibinfo{author}{\bibfnamefont{M.}~\bibnamefont{Ali}},
  \bibinfo{author}{\bibfnamefont{J.}~\bibnamefont{van~den Brink}},
  \bibinfo{author}{\bibfnamefont{M.}~\bibnamefont{Hoesch}},
  \bibinfo{author}{\bibfnamefont{A.}~\bibnamefont{Fedorov}},
  \bibnamefont{et~al.}, \bibinfo{journal}{Nature Communications}
  \textbf{\bibinfo{volume}{10}}, \bibinfo{pages}{3424} (\bibinfo{year}{2019}),
  \urlprefix\url{https://doi.org/10.1038/s41467-019-11393-5}.

\bibitem[{\citenamefont{Zhang et~al.}(2019)\citenamefont{Zhang, Shi, Zhu, Xing,
  Zhang, and Wang}}]{ZhangPRL}
\bibinfo{author}{\bibfnamefont{D.}~\bibnamefont{Zhang}},
  \bibinfo{author}{\bibfnamefont{M.}~\bibnamefont{Shi}},
  \bibinfo{author}{\bibfnamefont{T.}~\bibnamefont{Zhu}},
  \bibinfo{author}{\bibfnamefont{D.}~\bibnamefont{Xing}},
  \bibinfo{author}{\bibfnamefont{H.}~\bibnamefont{Zhang}}, \bibnamefont{and}
  \bibinfo{author}{\bibfnamefont{J.}~\bibnamefont{Wang}},
  \bibinfo{journal}{Phys. Rev. Lett.} \textbf{\bibinfo{volume}{122}},
  \bibinfo{pages}{206401} (\bibinfo{year}{2019}),
  \urlprefix\url{https://link.aps.org/doi/10.1103/PhysRevLett.122.206401}.

\bibitem[{\citenamefont{Rienks et~al.}(2019)\citenamefont{Rienks, Wimmer,
  S{\'a}nchez-Barriga, Caha, Mandal, R{\r u}{\v z}i{\v c}ka, Ney, Steiner,
  Volobuev, Groiss et~al.}}]{RienksNature}
\bibinfo{author}{\bibfnamefont{E.~D.~L.} \bibnamefont{Rienks}},
  \bibinfo{author}{\bibfnamefont{S.}~\bibnamefont{Wimmer}},
  \bibinfo{author}{\bibfnamefont{J.}~\bibnamefont{S{\'a}nchez-Barriga}},
  \bibinfo{author}{\bibfnamefont{O.}~\bibnamefont{Caha}},
  \bibinfo{author}{\bibfnamefont{P.~S.} \bibnamefont{Mandal}},
  \bibinfo{author}{\bibfnamefont{J.}~\bibnamefont{R{\r u}{\v z}i{\v c}ka}},
  \bibinfo{author}{\bibfnamefont{A.}~\bibnamefont{Ney}},
  \bibinfo{author}{\bibfnamefont{H.}~\bibnamefont{Steiner}},
  \bibinfo{author}{\bibfnamefont{V.~V.} \bibnamefont{Volobuev}},
  \bibinfo{author}{\bibfnamefont{H.}~\bibnamefont{Groiss}},
  \bibnamefont{et~al.}, \bibinfo{journal}{Nature}
  \textbf{\bibinfo{volume}{576}}, \bibinfo{pages}{423} (\bibinfo{year}{2019}),
  \urlprefix\url{https://doi.org/10.1038/s41586-019-1826-7}.

\bibitem[{\citenamefont{Li et~al.}(2019{\natexlab{a}})\citenamefont{Li, Gao,
  Duan, Xu, Zhu, Tian, Gao, Fan, Rao, Huang et~al.}}]{LiCondmat}
\bibinfo{author}{\bibfnamefont{H.}~\bibnamefont{Li}},
  \bibinfo{author}{\bibfnamefont{S.-Y.} \bibnamefont{Gao}},
  \bibinfo{author}{\bibfnamefont{S.-F.} \bibnamefont{Duan}},
  \bibinfo{author}{\bibfnamefont{Y.-F.} \bibnamefont{Xu}},
  \bibinfo{author}{\bibfnamefont{K.-J.} \bibnamefont{Zhu}},
  \bibinfo{author}{\bibfnamefont{S.-J.} \bibnamefont{Tian}},
  \bibinfo{author}{\bibfnamefont{J.-C.} \bibnamefont{Gao}},
  \bibinfo{author}{\bibfnamefont{W.-H.} \bibnamefont{Fan}},
  \bibinfo{author}{\bibfnamefont{Z.-C.} \bibnamefont{Rao}},
  \bibinfo{author}{\bibfnamefont{J.-R.} \bibnamefont{Huang}},
  \bibnamefont{et~al.}, \bibinfo{journal}{Phys. Rev. X}
  \textbf{\bibinfo{volume}{9}}, \bibinfo{pages}{041039}
  (\bibinfo{year}{2019}{\natexlab{a}}),
  \urlprefix\url{https://link.aps.org/doi/10.1103/PhysRevX.9.041039}.

\bibitem[{\citenamefont{Li et~al.}(2019{\natexlab{b}})\citenamefont{Li, Li, Du,
  Wang, Gu, Zhang, He, Duan, and Xu}}]{LiSciAdv}
\bibinfo{author}{\bibfnamefont{J.}~\bibnamefont{Li}},
  \bibinfo{author}{\bibfnamefont{Y.}~\bibnamefont{Li}},
  \bibinfo{author}{\bibfnamefont{S.}~\bibnamefont{Du}},
  \bibinfo{author}{\bibfnamefont{Z.}~\bibnamefont{Wang}},
  \bibinfo{author}{\bibfnamefont{B.-L.} \bibnamefont{Gu}},
  \bibinfo{author}{\bibfnamefont{S.-C.} \bibnamefont{Zhang}},
  \bibinfo{author}{\bibfnamefont{K.}~\bibnamefont{He}},
  \bibinfo{author}{\bibfnamefont{W.}~\bibnamefont{Duan}}, \bibnamefont{and}
  \bibinfo{author}{\bibfnamefont{Y.}~\bibnamefont{Xu}},
  \bibinfo{journal}{Science Advances} \textbf{\bibinfo{volume}{5}}
  (\bibinfo{year}{2019}{\natexlab{b}}),
  \eprint{https://advances.sciencemag.org/content/5/6/eaaw5685.full.pdf},
  \urlprefix\url{https://advances.sciencemag.org/content/5/6/eaaw5685}.

\bibitem[{\citenamefont{Hu et~al.}(2020)\citenamefont{Hu, Gordon, Liu, Liu,
  Zhou, Hao, Narayan, Emmanouilidou, Sun, Liu et~al.}}]{huCondmat}
\bibinfo{author}{\bibfnamefont{C.}~\bibnamefont{Hu}},
  \bibinfo{author}{\bibfnamefont{K.~N.} \bibnamefont{Gordon}},
  \bibinfo{author}{\bibfnamefont{P.}~\bibnamefont{Liu}},
  \bibinfo{author}{\bibfnamefont{J.}~\bibnamefont{Liu}},
  \bibinfo{author}{\bibfnamefont{X.}~\bibnamefont{Zhou}},
  \bibinfo{author}{\bibfnamefont{P.}~\bibnamefont{Hao}},
  \bibinfo{author}{\bibfnamefont{D.}~\bibnamefont{Narayan}},
  \bibinfo{author}{\bibfnamefont{E.}~\bibnamefont{Emmanouilidou}},
  \bibinfo{author}{\bibfnamefont{H.}~\bibnamefont{Sun}},
  \bibinfo{author}{\bibfnamefont{Y.}~\bibnamefont{Liu}}, \bibnamefont{et~al.},
  \bibinfo{journal}{Nature Communications} \textbf{\bibinfo{volume}{11}},
  \bibinfo{pages}{97} (\bibinfo{year}{2020}),
  \urlprefix\url{https://doi.org/10.1038/s41467-019-13814-x}.

\bibitem[{\citenamefont{Ma et~al.}(2019)\citenamefont{Ma, Nie, Yi, Jandke,
  Shang, Yao, Naamneh, Yan, Sun, Chikina et~al.}}]{MaSciAdv}
\bibinfo{author}{\bibfnamefont{J.-Z.} \bibnamefont{Ma}},
  \bibinfo{author}{\bibfnamefont{S.~M.} \bibnamefont{Nie}},
  \bibinfo{author}{\bibfnamefont{C.~J.} \bibnamefont{Yi}},
  \bibinfo{author}{\bibfnamefont{J.}~\bibnamefont{Jandke}},
  \bibinfo{author}{\bibfnamefont{T.}~\bibnamefont{Shang}},
  \bibinfo{author}{\bibfnamefont{M.~Y.} \bibnamefont{Yao}},
  \bibinfo{author}{\bibfnamefont{M.}~\bibnamefont{Naamneh}},
  \bibinfo{author}{\bibfnamefont{L.~Q.} \bibnamefont{Yan}},
  \bibinfo{author}{\bibfnamefont{Y.}~\bibnamefont{Sun}},
  \bibinfo{author}{\bibfnamefont{A.}~\bibnamefont{Chikina}},
  \bibnamefont{et~al.}, \bibinfo{journal}{Science Advances}
  \textbf{\bibinfo{volume}{5}} (\bibinfo{year}{2019}),
  \eprint{https://advances.sciencemag.org/content/5/7/eaaw4718.full.pdf},
  \urlprefix\url{https://advances.sciencemag.org/content/5/7/eaaw4718}.

\bibitem[{\citenamefont{Ye et~al.}(2018)\citenamefont{Ye, Kang, Liu, von Cube,
  Wicker, Suzuki, Jozwiak, Bostwick, Rotenberg, Bell et~al.}}]{YeNature}
\bibinfo{author}{\bibfnamefont{L.}~\bibnamefont{Ye}},
  \bibinfo{author}{\bibfnamefont{M.}~\bibnamefont{Kang}},
  \bibinfo{author}{\bibfnamefont{J.}~\bibnamefont{Liu}},
  \bibinfo{author}{\bibfnamefont{F.}~\bibnamefont{von Cube}},
  \bibinfo{author}{\bibfnamefont{C.~R.} \bibnamefont{Wicker}},
  \bibinfo{author}{\bibfnamefont{T.}~\bibnamefont{Suzuki}},
  \bibinfo{author}{\bibfnamefont{C.}~\bibnamefont{Jozwiak}},
  \bibinfo{author}{\bibfnamefont{A.}~\bibnamefont{Bostwick}},
  \bibinfo{author}{\bibfnamefont{E.}~\bibnamefont{Rotenberg}},
  \bibinfo{author}{\bibfnamefont{D.~C.} \bibnamefont{Bell}},
  \bibnamefont{et~al.}, \bibinfo{journal}{Nature}
  \textbf{\bibinfo{volume}{555}}, \bibinfo{pages}{638} (\bibinfo{year}{2018}),
  \urlprefix\url{https://doi.org/10.1038/nature25987}.

\bibitem[{\citenamefont{Yin et~al.}(2018)\citenamefont{Yin, Zhang, Li, Jiang,
  Chang, Zhang, Lian, Xiang, Belopolski, Zheng et~al.}}]{YinNature}
\bibinfo{author}{\bibfnamefont{J.-X.} \bibnamefont{Yin}},
  \bibinfo{author}{\bibfnamefont{S.~S.} \bibnamefont{Zhang}},
  \bibinfo{author}{\bibfnamefont{H.}~\bibnamefont{Li}},
  \bibinfo{author}{\bibfnamefont{K.}~\bibnamefont{Jiang}},
  \bibinfo{author}{\bibfnamefont{G.}~\bibnamefont{Chang}},
  \bibinfo{author}{\bibfnamefont{B.}~\bibnamefont{Zhang}},
  \bibinfo{author}{\bibfnamefont{B.}~\bibnamefont{Lian}},
  \bibinfo{author}{\bibfnamefont{C.}~\bibnamefont{Xiang}},
  \bibinfo{author}{\bibfnamefont{I.}~\bibnamefont{Belopolski}},
  \bibinfo{author}{\bibfnamefont{H.}~\bibnamefont{Zheng}},
  \bibnamefont{et~al.}, \bibinfo{journal}{Nature}
  \textbf{\bibinfo{volume}{562}}, \bibinfo{pages}{91} (\bibinfo{year}{2018}),
  \urlprefix\url{https://doi.org/10.1038/s41586-018-0502-7}.

\bibitem[{\citenamefont{Kang et~al.}(2019)\citenamefont{Kang, Ye, Fang, You,
  Levitan, Han, Facio, Jozwiak, Bostwick, Rotenberg et~al.}}]{KangNMat}
\bibinfo{author}{\bibfnamefont{M.}~\bibnamefont{Kang}},
  \bibinfo{author}{\bibfnamefont{L.}~\bibnamefont{Ye}},
  \bibinfo{author}{\bibfnamefont{S.}~\bibnamefont{Fang}},
  \bibinfo{author}{\bibfnamefont{J.-S.} \bibnamefont{You}},
  \bibinfo{author}{\bibfnamefont{A.}~\bibnamefont{Levitan}},
  \bibinfo{author}{\bibfnamefont{M.}~\bibnamefont{Han}},
  \bibinfo{author}{\bibfnamefont{J.~I.} \bibnamefont{Facio}},
  \bibinfo{author}{\bibfnamefont{C.}~\bibnamefont{Jozwiak}},
  \bibinfo{author}{\bibfnamefont{A.}~\bibnamefont{Bostwick}},
  \bibinfo{author}{\bibfnamefont{E.}~\bibnamefont{Rotenberg}},
  \bibnamefont{et~al.}, \bibinfo{journal}{Nature Materials}
  (\bibinfo{year}{2019}),
  \urlprefix\url{https://doi.org/10.1038/s41563-019-0531-0}.

\bibitem[{\citenamefont{Arguilla et~al.}(2017)\citenamefont{Arguilla, Cultrara,
  Baum, Jiang, Ross, and Goldberger}}]{Arguilla}
\bibinfo{author}{\bibfnamefont{M.~Q.} \bibnamefont{Arguilla}},
  \bibinfo{author}{\bibfnamefont{N.~D.} \bibnamefont{Cultrara}},
  \bibinfo{author}{\bibfnamefont{Z.~J.} \bibnamefont{Baum}},
  \bibinfo{author}{\bibfnamefont{S.}~\bibnamefont{Jiang}},
  \bibinfo{author}{\bibfnamefont{R.~D.} \bibnamefont{Ross}}, \bibnamefont{and}
  \bibinfo{author}{\bibfnamefont{J.~E.} \bibnamefont{Goldberger}},
  \bibinfo{journal}{Inorg. Chem. Front.} \textbf{\bibinfo{volume}{4}},
  \bibinfo{pages}{378} (\bibinfo{year}{2017}),
  \urlprefix\url{http://dx.doi.org/10.1039/C6QI00476H}.

\bibitem[{\citenamefont{Sakagami et~al.}(2018)\citenamefont{Sakagami, Goto,
  Mizuguchi, Matoba, and Kamihara}}]{Sakagami}
\bibinfo{author}{\bibfnamefont{R.}~\bibnamefont{Sakagami}},
  \bibinfo{author}{\bibfnamefont{Y.}~\bibnamefont{Goto}},
  \bibinfo{author}{\bibfnamefont{Y.}~\bibnamefont{Mizuguchi}},
  \bibinfo{author}{\bibfnamefont{M.}~\bibnamefont{Matoba}}, \bibnamefont{and}
  \bibinfo{author}{\bibfnamefont{Y.}~\bibnamefont{Kamihara}},
  \bibinfo{journal}{Mater. Sci. Tech. Jpn.} \textbf{\bibinfo{volume}{55}},
  \bibinfo{pages}{72} (\bibinfo{year}{2018}).

\bibitem[{\citenamefont{Eley et~al.}(2012)\citenamefont{Eley, Gopalakrishnan,
  Goldbart, and Mason}}]{EleyNPhys}
\bibinfo{author}{\bibfnamefont{S.}~\bibnamefont{Eley}},
  \bibinfo{author}{\bibfnamefont{S.}~\bibnamefont{Gopalakrishnan}},
  \bibinfo{author}{\bibfnamefont{P.~M.} \bibnamefont{Goldbart}},
  \bibnamefont{and} \bibinfo{author}{\bibfnamefont{N.}~\bibnamefont{Mason}},
  \bibinfo{journal}{Nature Physics} \textbf{\bibinfo{volume}{8}},
  \bibinfo{pages}{59} (\bibinfo{year}{2012}),
  \urlprefix\url{https://doi.org/10.1038/nphys2154}.

\bibitem[{\citenamefont{Chen et~al.}(2018)\citenamefont{Chen, Swartz, Yoon,
  Inoue, Merz, Lu, Xie, Yuan, Hikita, Raghu et~al.}}]{ChenNCom}
\bibinfo{author}{\bibfnamefont{Z.}~\bibnamefont{Chen}},
  \bibinfo{author}{\bibfnamefont{A.~G.} \bibnamefont{Swartz}},
  \bibinfo{author}{\bibfnamefont{H.}~\bibnamefont{Yoon}},
  \bibinfo{author}{\bibfnamefont{H.}~\bibnamefont{Inoue}},
  \bibinfo{author}{\bibfnamefont{T.~A.} \bibnamefont{Merz}},
  \bibinfo{author}{\bibfnamefont{D.}~\bibnamefont{Lu}},
  \bibinfo{author}{\bibfnamefont{Y.}~\bibnamefont{Xie}},
  \bibinfo{author}{\bibfnamefont{H.}~\bibnamefont{Yuan}},
  \bibinfo{author}{\bibfnamefont{Y.}~\bibnamefont{Hikita}},
  \bibinfo{author}{\bibfnamefont{S.}~\bibnamefont{Raghu}},
  \bibnamefont{et~al.}, \bibinfo{journal}{Nature Communications}
  \textbf{\bibinfo{volume}{9}}, \bibinfo{pages}{4008} (\bibinfo{year}{2018}),
  \urlprefix\url{https://doi.org/10.1038/s41467-018-06444-2}.

\bibitem[{\citenamefont{Goto et~al.}(2017)\citenamefont{Goto, Yamada, Matsuda,
  Aoki, and Mizuguchi}}]{GotoJPSJ}
\bibinfo{author}{\bibfnamefont{Y.}~\bibnamefont{Goto}},
  \bibinfo{author}{\bibfnamefont{A.}~\bibnamefont{Yamada}},
  \bibinfo{author}{\bibfnamefont{T.~D.} \bibnamefont{Matsuda}},
  \bibinfo{author}{\bibfnamefont{Y.}~\bibnamefont{Aoki}}, \bibnamefont{and}
  \bibinfo{author}{\bibfnamefont{Y.}~\bibnamefont{Mizuguchi}},
  \bibinfo{journal}{Journal of the Physical Society of Japan}
  \textbf{\bibinfo{volume}{86}}, \bibinfo{pages}{123701}
  (\bibinfo{year}{2017}), \eprint{https://doi.org/10.7566/JPSJ.86.123701},
  \urlprefix\url{https://doi.org/10.7566/JPSJ.86.123701}.

\bibitem[{\citenamefont{Ishihara et~al.}(2018)\citenamefont{Ishihara, Takenaka,
  Miao, Tanaka, Mizukami, Usui, Kuroki, Konczykowski, Goto, Mizuguchi
  et~al.}}]{IshiharaPRB}
\bibinfo{author}{\bibfnamefont{K.}~\bibnamefont{Ishihara}},
  \bibinfo{author}{\bibfnamefont{T.}~\bibnamefont{Takenaka}},
  \bibinfo{author}{\bibfnamefont{Y.}~\bibnamefont{Miao}},
  \bibinfo{author}{\bibfnamefont{O.}~\bibnamefont{Tanaka}},
  \bibinfo{author}{\bibfnamefont{Y.}~\bibnamefont{Mizukami}},
  \bibinfo{author}{\bibfnamefont{H.}~\bibnamefont{Usui}},
  \bibinfo{author}{\bibfnamefont{K.}~\bibnamefont{Kuroki}},
  \bibinfo{author}{\bibfnamefont{M.}~\bibnamefont{Konczykowski}},
  \bibinfo{author}{\bibfnamefont{Y.}~\bibnamefont{Goto}},
  \bibinfo{author}{\bibfnamefont{Y.}~\bibnamefont{Mizuguchi}},
  \bibnamefont{et~al.}, \bibinfo{journal}{Phys. Rev. B}
  \textbf{\bibinfo{volume}{98}}, \bibinfo{pages}{020503}
  (\bibinfo{year}{2018}),
  \urlprefix\url{https://link.aps.org/doi/10.1103/PhysRevB.98.020503}.

\bibitem[{\citenamefont{Gui et~al.}(2019)\citenamefont{Gui, Pletikosic, Cao,
  Tien, Xu, Zhong, Wang, Chang, Jia, Valla et~al.}}]{GuiACS}
\bibinfo{author}{\bibfnamefont{X.}~\bibnamefont{Gui}},
  \bibinfo{author}{\bibfnamefont{I.}~\bibnamefont{Pletikosic}},
  \bibinfo{author}{\bibfnamefont{H.}~\bibnamefont{Cao}},
  \bibinfo{author}{\bibfnamefont{H.-J.} \bibnamefont{Tien}},
  \bibinfo{author}{\bibfnamefont{X.}~\bibnamefont{Xu}},
  \bibinfo{author}{\bibfnamefont{R.}~\bibnamefont{Zhong}},
  \bibinfo{author}{\bibfnamefont{G.}~\bibnamefont{Wang}},
  \bibinfo{author}{\bibfnamefont{T.-R.} \bibnamefont{Chang}},
  \bibinfo{author}{\bibfnamefont{S.}~\bibnamefont{Jia}},
  \bibinfo{author}{\bibfnamefont{T.}~\bibnamefont{Valla}},
  \bibnamefont{et~al.}, \bibinfo{journal}{ACS Central Science}
  \textbf{\bibinfo{volume}{5}}, \bibinfo{pages}{900} (\bibinfo{year}{2019}),
  \urlprefix\url{https://doi.org/10.1021/acscentsci.9b00202}.

\bibitem[{\citenamefont{Ota et~al.}(2017)\citenamefont{Ota, Okazaki, Yamamoto,
  Yamamoto, Watanabe, Chen, Nagao, Watauchi, Tanaka, Takano et~al.}}]{OtaPRL}
\bibinfo{author}{\bibfnamefont{Y.}~\bibnamefont{Ota}},
  \bibinfo{author}{\bibfnamefont{K.}~\bibnamefont{Okazaki}},
  \bibinfo{author}{\bibfnamefont{H.~Q.} \bibnamefont{Yamamoto}},
  \bibinfo{author}{\bibfnamefont{T.}~\bibnamefont{Yamamoto}},
  \bibinfo{author}{\bibfnamefont{S.}~\bibnamefont{Watanabe}},
  \bibinfo{author}{\bibfnamefont{C.}~\bibnamefont{Chen}},
  \bibinfo{author}{\bibfnamefont{M.}~\bibnamefont{Nagao}},
  \bibinfo{author}{\bibfnamefont{S.}~\bibnamefont{Watauchi}},
  \bibinfo{author}{\bibfnamefont{I.}~\bibnamefont{Tanaka}},
  \bibinfo{author}{\bibfnamefont{Y.}~\bibnamefont{Takano}},
  \bibnamefont{et~al.}, \bibinfo{journal}{Phys. Rev. Lett.}
  \textbf{\bibinfo{volume}{118}}, \bibinfo{pages}{167002}
  (\bibinfo{year}{2017}),
  \urlprefix\url{https://link.aps.org/doi/10.1103/PhysRevLett.118.167002}.

\bibitem[{\citenamefont{Saito et~al.}(2015)\citenamefont{Saito, Nakamura,
  Bahramy, Kohama, Ye, Kasahara, Nakagawa, Onga, Tokunaga, Nojima
  et~al.}}]{SaitoNPhys}
\bibinfo{author}{\bibfnamefont{Y.}~\bibnamefont{Saito}},
  \bibinfo{author}{\bibfnamefont{Y.}~\bibnamefont{Nakamura}},
  \bibinfo{author}{\bibfnamefont{M.~S.} \bibnamefont{Bahramy}},
  \bibinfo{author}{\bibfnamefont{Y.}~\bibnamefont{Kohama}},
  \bibinfo{author}{\bibfnamefont{J.}~\bibnamefont{Ye}},
  \bibinfo{author}{\bibfnamefont{Y.}~\bibnamefont{Kasahara}},
  \bibinfo{author}{\bibfnamefont{Y.}~\bibnamefont{Nakagawa}},
  \bibinfo{author}{\bibfnamefont{M.}~\bibnamefont{Onga}},
  \bibinfo{author}{\bibfnamefont{M.}~\bibnamefont{Tokunaga}},
  \bibinfo{author}{\bibfnamefont{T.}~\bibnamefont{Nojima}},
  \bibnamefont{et~al.}, \bibinfo{journal}{Nature Physics}
  \textbf{\bibinfo{volume}{12}}, \bibinfo{pages}{144 EP }
  (\bibinfo{year}{2015}), \urlprefix\url{https://doi.org/10.1038/nphys3580}.

\bibitem[{\citenamefont{Kawabata et~al.}(2010)\citenamefont{Kawabata, Asano,
  Tanaka, Golubov, and Kashiwaya}}]{KawabataPRL}
\bibinfo{author}{\bibfnamefont{S.}~\bibnamefont{Kawabata}},
  \bibinfo{author}{\bibfnamefont{Y.}~\bibnamefont{Asano}},
  \bibinfo{author}{\bibfnamefont{Y.}~\bibnamefont{Tanaka}},
  \bibinfo{author}{\bibfnamefont{A.~A.} \bibnamefont{Golubov}},
  \bibnamefont{and}
  \bibinfo{author}{\bibfnamefont{S.}~\bibnamefont{Kashiwaya}},
  \bibinfo{journal}{Phys. Rev. Lett.} \textbf{\bibinfo{volume}{104}},
  \bibinfo{pages}{117002} (\bibinfo{year}{2010}),
  \urlprefix\url{https://link.aps.org/doi/10.1103/PhysRevLett.104.117002}.

\bibitem[{\citenamefont{Kawabata and Asano}(2010)}]{KawabataLTP}
\bibinfo{author}{\bibfnamefont{S.}~\bibnamefont{Kawabata}} \bibnamefont{and}
  \bibinfo{author}{\bibfnamefont{Y.}~\bibnamefont{Asano}},
  \bibinfo{journal}{Low Temperature Physics} \textbf{\bibinfo{volume}{36}},
  \bibinfo{pages}{915} (\bibinfo{year}{2010}),
  \eprint{https://doi.org/10.1063/1.3515524},
  \urlprefix\url{https://doi.org/10.1063/1.3515524}.

\bibitem[{\citenamefont{Baltz et~al.}(2018)\citenamefont{Baltz, Manchon, Tsoi,
  Moriyama, Ono, and Tserkovnyak}}]{BaltzRMP}
\bibinfo{author}{\bibfnamefont{V.}~\bibnamefont{Baltz}},
  \bibinfo{author}{\bibfnamefont{A.}~\bibnamefont{Manchon}},
  \bibinfo{author}{\bibfnamefont{M.}~\bibnamefont{Tsoi}},
  \bibinfo{author}{\bibfnamefont{T.}~\bibnamefont{Moriyama}},
  \bibinfo{author}{\bibfnamefont{T.}~\bibnamefont{Ono}}, \bibnamefont{and}
  \bibinfo{author}{\bibfnamefont{Y.}~\bibnamefont{Tserkovnyak}},
  \bibinfo{journal}{Rev. Mod. Phys.} \textbf{\bibinfo{volume}{90}},
  \bibinfo{pages}{015005} (\bibinfo{year}{2018}),
  \urlprefix\url{https://link.aps.org/doi/10.1103/RevModPhys.90.015005}.

\bibitem[{\citenamefont{N{\v e}mec et~al.}(2018)\citenamefont{N{\v e}mec,
  Fiebig, Kampfrath, and Kimel}}]{NemecNPhys}
\bibinfo{author}{\bibfnamefont{P.}~\bibnamefont{N{\v e}mec}},
  \bibinfo{author}{\bibfnamefont{M.}~\bibnamefont{Fiebig}},
  \bibinfo{author}{\bibfnamefont{T.}~\bibnamefont{Kampfrath}},
  \bibnamefont{and} \bibinfo{author}{\bibfnamefont{A.~V.} \bibnamefont{Kimel}},
  \bibinfo{journal}{Nature Physics} \textbf{\bibinfo{volume}{14}},
  \bibinfo{pages}{229} (\bibinfo{year}{2018}),
  \urlprefix\url{https://doi.org/10.1038/s41567-018-0051-x}.

\bibitem[{\citenamefont{Nakao et~al.}(2014)\citenamefont{Nakao, Yamasaki,
  Okamoto, Sudayama, Takahashi, Kobayashi, Kumai, and Murakami}}]{NakaoJPSC}
\bibinfo{author}{\bibfnamefont{H.}~\bibnamefont{Nakao}},
  \bibinfo{author}{\bibfnamefont{Y.}~\bibnamefont{Yamasaki}},
  \bibinfo{author}{\bibfnamefont{J.}~\bibnamefont{Okamoto}},
  \bibinfo{author}{\bibfnamefont{T.}~\bibnamefont{Sudayama}},
  \bibinfo{author}{\bibfnamefont{Y.}~\bibnamefont{Takahashi}},
  \bibinfo{author}{\bibfnamefont{K.}~\bibnamefont{Kobayashi}},
  \bibinfo{author}{\bibfnamefont{R.}~\bibnamefont{Kumai}}, \bibnamefont{and}
  \bibinfo{author}{\bibfnamefont{Y.}~\bibnamefont{Murakami}},
  \bibinfo{journal}{Journal of Physics: Conference Series}
  \textbf{\bibinfo{volume}{502}}, \bibinfo{pages}{012015}
  (\bibinfo{year}{2014}),
  \urlprefix\url{https://doi.org/10.1088%2F1742-6596%2F502%2F1%2F012015}.

\bibitem[{\citenamefont{Katakura et~al.}(2010)\citenamefont{Katakura, Tokunaga,
  Matsuo, Kawaguchi, Kindo, Hitomi, Akahoshi, and Kuwahara}}]{TokunagaRSI}
\bibinfo{author}{\bibfnamefont{I.}~\bibnamefont{Katakura}},
  \bibinfo{author}{\bibfnamefont{M.}~\bibnamefont{Tokunaga}},
  \bibinfo{author}{\bibfnamefont{A.}~\bibnamefont{Matsuo}},
  \bibinfo{author}{\bibfnamefont{K.}~\bibnamefont{Kawaguchi}},
  \bibinfo{author}{\bibfnamefont{K.}~\bibnamefont{Kindo}},
  \bibinfo{author}{\bibfnamefont{M.}~\bibnamefont{Hitomi}},
  \bibinfo{author}{\bibfnamefont{D.}~\bibnamefont{Akahoshi}}, \bibnamefont{and}
  \bibinfo{author}{\bibfnamefont{H.}~\bibnamefont{Kuwahara}},
  \bibinfo{journal}{Review of Scientific Instruments}
  \textbf{\bibinfo{volume}{81}}, \bibinfo{pages}{043701}
  (\bibinfo{year}{2010}), \eprint{https://doi.org/10.1063/1.3359954},
  \urlprefix\url{https://doi.org/10.1063/1.3359954}.

\bibitem[{\citenamefont{Kihara et~al.}(2013)\citenamefont{Kihara, Kohama,
  Hashimoto, Katsumoto, and Tokunaga}}]{TokunagaRSI2}
\bibinfo{author}{\bibfnamefont{T.}~\bibnamefont{Kihara}},
  \bibinfo{author}{\bibfnamefont{Y.}~\bibnamefont{Kohama}},
  \bibinfo{author}{\bibfnamefont{Y.}~\bibnamefont{Hashimoto}},
  \bibinfo{author}{\bibfnamefont{S.}~\bibnamefont{Katsumoto}},
  \bibnamefont{and} \bibinfo{author}{\bibfnamefont{M.}~\bibnamefont{Tokunaga}},
  \bibinfo{journal}{Review of Scientific Instruments}
  \textbf{\bibinfo{volume}{84}}, \bibinfo{pages}{074901}
  (\bibinfo{year}{2013}), \eprint{https://doi.org/10.1063/1.4811798},
  \urlprefix\url{https://doi.org/10.1063/1.4811798}.

\bibitem[{\citenamefont{Kresse and Hafner}(1993)}]{VASP}
\bibinfo{author}{\bibfnamefont{G.}~\bibnamefont{Kresse}} \bibnamefont{and}
  \bibinfo{author}{\bibfnamefont{J.}~\bibnamefont{Hafner}},
  \bibinfo{journal}{Phys. Rev. B} \textbf{\bibinfo{volume}{47}},
  \bibinfo{pages}{558} (\bibinfo{year}{1993}),
  \urlprefix\url{https://link.aps.org/doi/10.1103/PhysRevB.47.558}.

\bibitem[{\citenamefont{Bl\"ochl}(1994)}]{PAW}
\bibinfo{author}{\bibfnamefont{P.~E.} \bibnamefont{Bl\"ochl}},
  \bibinfo{journal}{Phys. Rev. B} \textbf{\bibinfo{volume}{50}},
  \bibinfo{pages}{17953} (\bibinfo{year}{1994}),
  \urlprefix\url{https://link.aps.org/doi/10.1103/PhysRevB.50.17953}.

\bibitem[{\citenamefont{Perdew et~al.}(1996)\citenamefont{Perdew, Burke, and
  Ernzerhof}}]{GGAPBE}
\bibinfo{author}{\bibfnamefont{J.~P.} \bibnamefont{Perdew}},
  \bibinfo{author}{\bibfnamefont{K.}~\bibnamefont{Burke}}, \bibnamefont{and}
  \bibinfo{author}{\bibfnamefont{M.}~\bibnamefont{Ernzerhof}},
  \bibinfo{journal}{Phys. Rev. Lett.} \textbf{\bibinfo{volume}{77}},
  \bibinfo{pages}{3865} (\bibinfo{year}{1996}),
  \urlprefix\url{https://link.aps.org/doi/10.1103/PhysRevLett.77.3865}.

\bibitem[{\citenamefont{Dudarev et~al.}(1998)\citenamefont{Dudarev, Botton,
  Savrasov, Humphreys, and Sutton}}]{DFTU}
\bibinfo{author}{\bibfnamefont{S.~L.} \bibnamefont{Dudarev}},
  \bibinfo{author}{\bibfnamefont{G.~A.} \bibnamefont{Botton}},
  \bibinfo{author}{\bibfnamefont{S.~Y.} \bibnamefont{Savrasov}},
  \bibinfo{author}{\bibfnamefont{C.~J.} \bibnamefont{Humphreys}},
  \bibnamefont{and} \bibinfo{author}{\bibfnamefont{A.~P.}
  \bibnamefont{Sutton}}, \bibinfo{journal}{Phys. Rev. B}
  \textbf{\bibinfo{volume}{57}}, \bibinfo{pages}{1505} (\bibinfo{year}{1998}),
  \urlprefix\url{https://link.aps.org/doi/10.1103/PhysRevB.57.1505}.

\bibitem[{\citenamefont{Mostofi et~al.}(2014)\citenamefont{Mostofi, Yates,
  Pizzi, Lee, Souza, Vanderbilt, and Marzari}}]{wannier90}
\bibinfo{author}{\bibfnamefont{A.~A.} \bibnamefont{Mostofi}},
  \bibinfo{author}{\bibfnamefont{J.~R.} \bibnamefont{Yates}},
  \bibinfo{author}{\bibfnamefont{G.}~\bibnamefont{Pizzi}},
  \bibinfo{author}{\bibfnamefont{Y.-S.} \bibnamefont{Lee}},
  \bibinfo{author}{\bibfnamefont{I.}~\bibnamefont{Souza}},
  \bibinfo{author}{\bibfnamefont{D.}~\bibnamefont{Vanderbilt}},
  \bibnamefont{and} \bibinfo{author}{\bibfnamefont{N.}~\bibnamefont{Marzari}},
  \bibinfo{journal}{Computer Physics Communications}
  \textbf{\bibinfo{volume}{185}}, \bibinfo{pages}{2309 }
  (\bibinfo{year}{2014}), ISSN \bibinfo{issn}{0010-4655},
  \urlprefix\url{http://www.sciencedirect.com/science/article/pii/S001046551400157X}.

\end{thebibliography}

\begin{figure}
\includegraphics[width=6.5in]{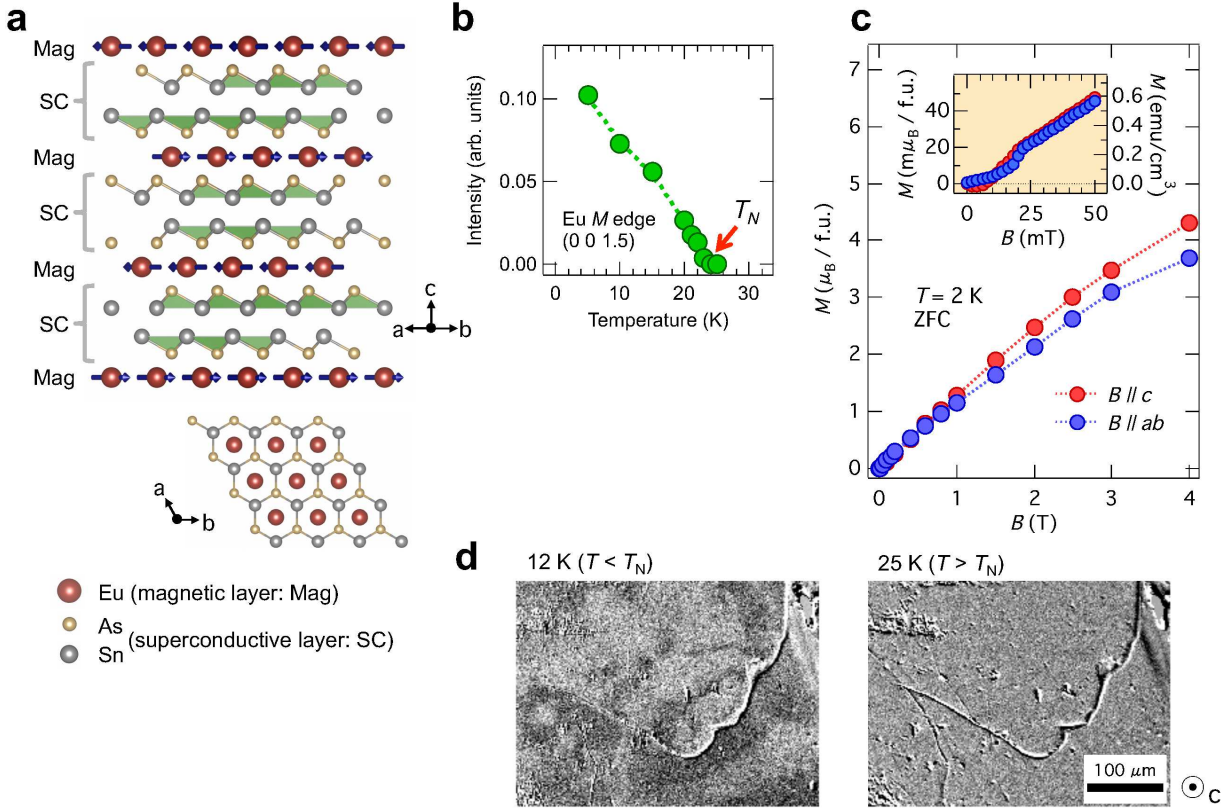}
\renewcommand{\baselinestretch}{1.1}
\caption{ {\bf Crystal structure and magnetic properties of  EuSn$_2$As$_2$.}
{\bf a,} The crystal structure of EuSn$_2$As$_2$ viewed parallel (top) and perpendicular (bottom) with respect to the $a$-$b$ plane.  This compound has a layered structure (so-called Zintle phase) with the space group $R\overline{3}m$, and the Sn-As-Eu-As-Sn blocks are separated by the van der Waals gap from each other \cite{Arguilla, Sakagami}. 
{\bf b,} The temperature dependence of the integrated intensity of (0 0 1.5) reflection obtained by the X-ray magnetic scattering at the $M_5$-edge of Eu, demonstrating that the antiferromagnetic ordering of the Eu moment occurs along the $c$-axis below $T_N \sim 23$~K (Figs. S1, 2). 
{\bf c,} The field dependence of the magnetization in EuSn$_2$As$_2$  at  $B \parallel c$ (red) and $B \parallel ab$ (blue) measured at 2 K after the zero-field cooling (ZFC). The inset zooms a region of small magnetic fields for the same data.
{\bf d,} Magneto-optical images for the surface of EuSn$_2$As$_2$ in the AFM phase (left, taken at 12 K) and the paramagnetic phase (right, 25 K). 
Magneto-optical images for the surface of EuSn$_2$As$_2$ in the AFM phase (left, taken at 12 K) and the paramagnetic phase (right, 25 K). The Kerr microscope was adjusted to become sensitive to the in-plane moment \cite{BorisenkoNComm}; a stronger intensity is exhibited by a brighter color. The patch pattern emerging in the AFM phase indicates the formation of  inhomogeneously distributed ferromagnetic domains.
}
\end{figure}

\begin{figure}
\includegraphics[width=6.5in]{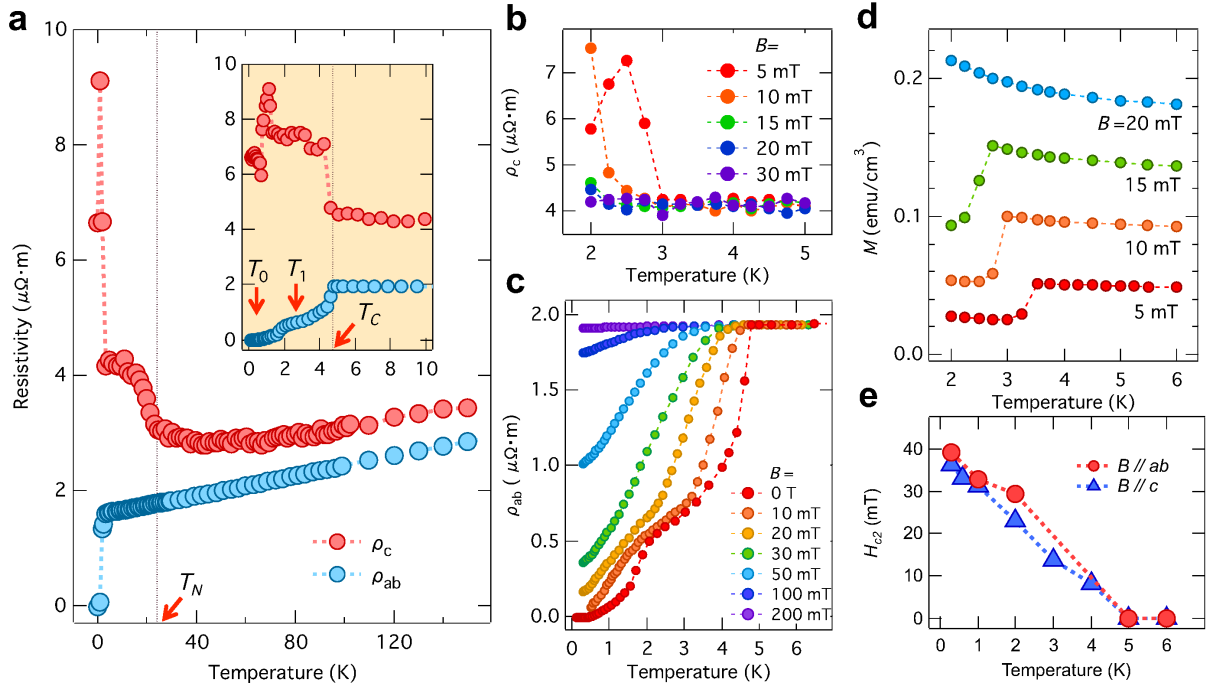}
\renewcommand{\baselinestretch}{1.1}
\caption{ {\bf Superconducting transport and magnetic properties of EuSn$_2$As$_2$.}
{\bf a,} The temperature dependence of the in-plane ($\rho_{ab}$) and out-of-plane ($\rho_{c}$) resistivities. The inset shows an enlarged view of the low temperature data. The critical ($T_c$) and the zero-resistivity ($T_0$) temperature are indicated by arrows. In addition, a shoulder-like feature yielding the two-step transition is also marked by an arrow at  $T_1$. 
{\bf b, c,} The variation of $\rho_{c}(T)$ and $\rho_{ab}(T)$ with different external magnetic fields, respectively. 
{\bf d,} The temperature dependence of the magnetization measured at $B \parallel ab$ by the field cooling.
{\bf e,} The temperature dependence of the critical magnetic field ($H_{c2}$) measured at $B \parallel ab$ (red circles) and $B \parallel c$ (blue triangles); each value was determined as $B$ at which $\rho_{ab}(B)$ reaches the mid-point of the transition (Fig. S 3).
}
\end{figure}

\begin{figure}
\includegraphics[width=5.7in]{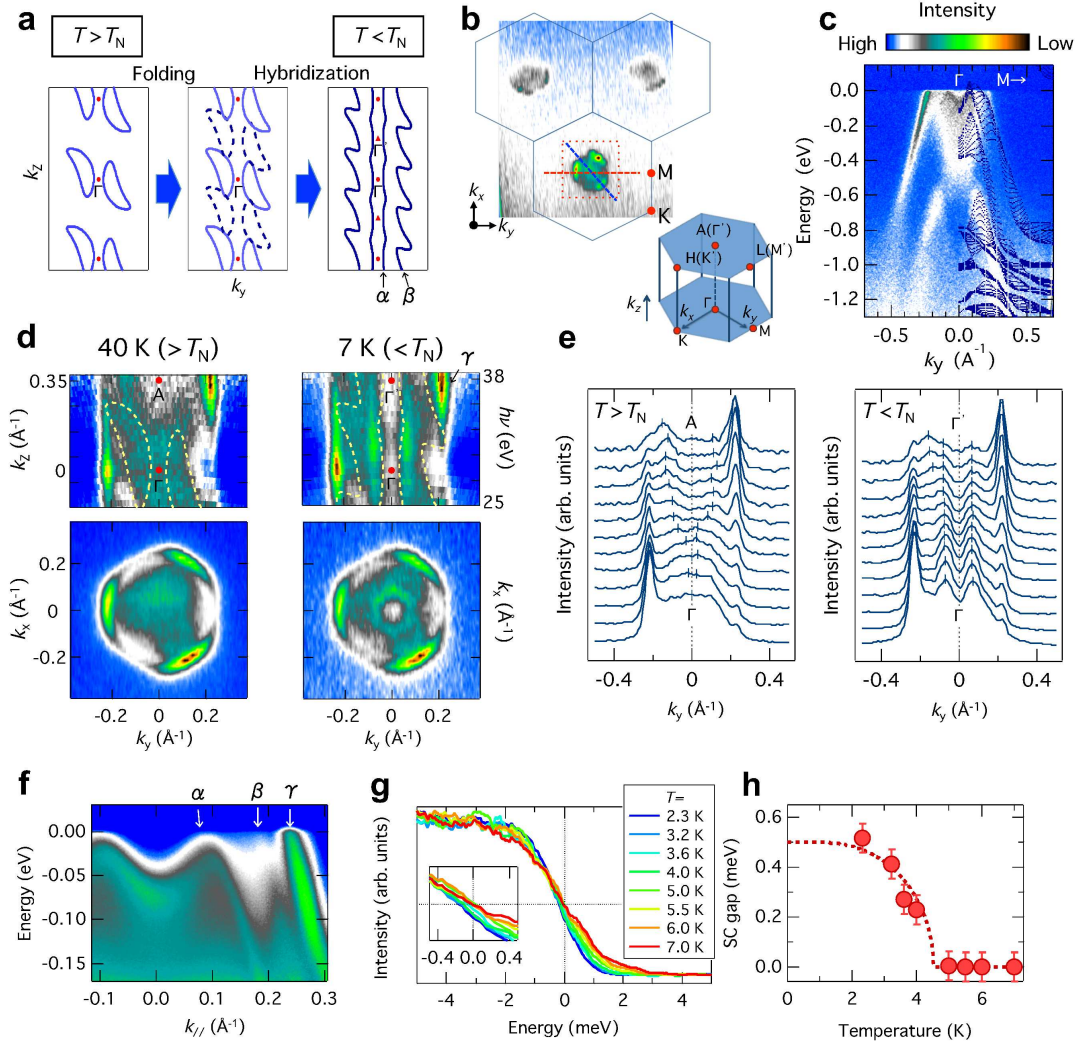}
\renewcommand{\baselinestretch}{1.1}
\caption{{\bf Band structure and superconducting gap measured by ARPES.}
{\bf a,} The temperature variation of Fermi surface along $k_z$ across $T_N$ expected from the DFT-based analysis (see also Fig. S4). The 2D and 3D Fermi surfaces in the AFM phase (right panel) are labeled by $\alpha$ and $\beta$, respectively.
{\bf b,} Fermi surface map along $k_y$-$k_x$ over multiple Brillouin zones measured at $T=7$~K and $h\nu = 70$~eV. 
{\bf c,} APRES band dispersion along the $\Gamma$-M line in {\bf b} (red dashed line); the DFT calculations are overlapped. 
{\bf d,} The top panels plot Fermi surface maps along $k_y$-$k_z$ measured at $k_x$=0 (red dashed line in {\bf b}) by changing photon energies above (left) and below (right) $T_N$. In the bottom, the Fermi surface maps along $k_y$-$k_x$ across $k_z$=0 ($h\nu$=28.5 eV) are plotted; momentum region of the measurements is indicated by a dashed square in {\bf b}. 
{\bf e,} Momentum distribution curves at 40 K (left) and 7 K (right) extracted from the top images in {\bf d}. 
{\bf f,} The dispersion map along the blue line in {\bf b} measured at 6 K by a laser-based APRES with high momentum and energy resolutions. Three bands of the surface state ($\gamma$) and bulk states ($\alpha$ and $\beta$) are labeled by  white arrows. 
{\bf g,} The energy distribution curves (EDCs) at $k_F$ of the $\beta$-band indicated by a white arrow in {\bf f}. 
{\bf h,} Temperature dependence of the superconducting gap estimated by fitting the BCS spectral function to EDCs in {\bf g}. The dashed curve shows the temperature evolution of energy gap for a theoretical BCS superconductor with $T_c$ of 4.8 K, same as that of our sample.
}
\end{figure}

\begin{figure}
\includegraphics[width=6.5in]{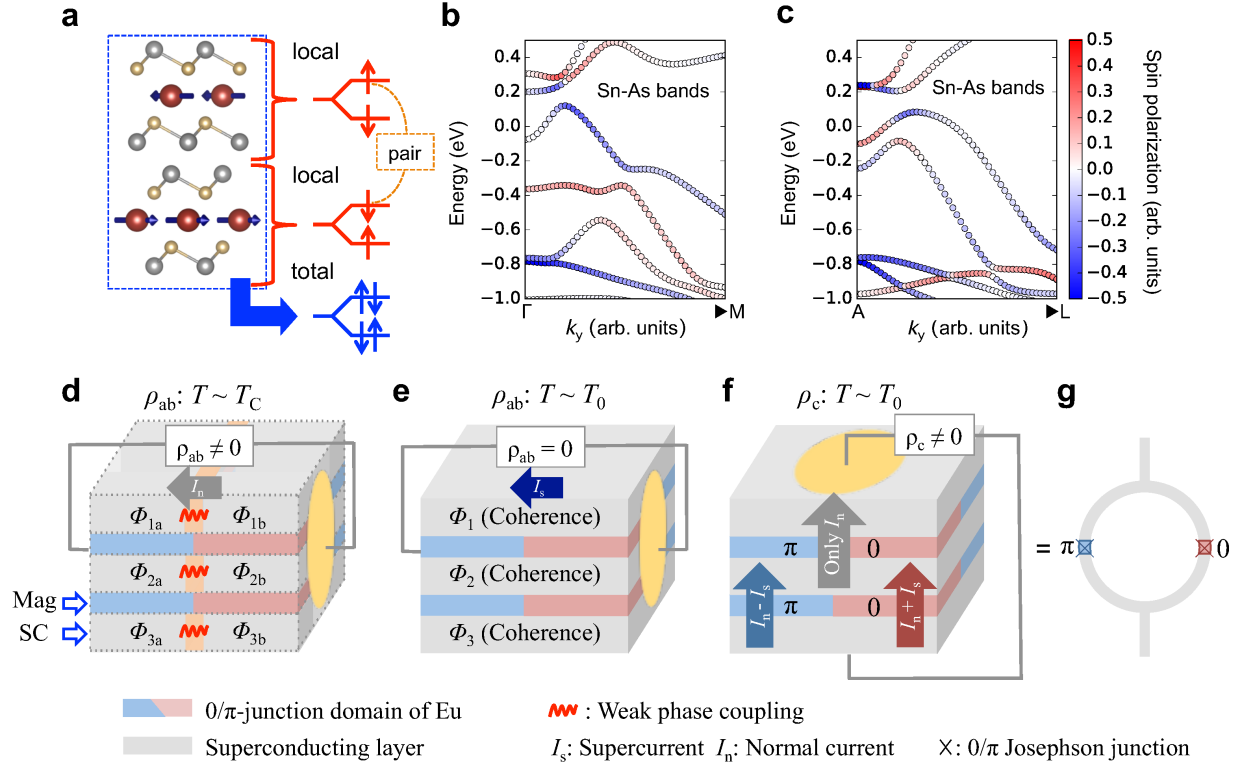}
\renewcommand{\baselinestretch}{1.1}
\caption{{\bf Spin polarization of local electronic states and a proposed mechanism of superconducting transport in EuSn$_2$As$_2$.} 
{\bf a,} A schematic illustration of the spin-splitting of conductive electrons in the Sn-As sheets adjacent to the ferromagnetic Eu layer. The up- and down-states are flipped in energy for the opposite direction of the ferromagnetism, which cancels the total spin-polarization in bulk.  
{\bf b, c,} The spin-polarized bands dominated by the Sn and As orbitals along $\Gamma$-M and A-L, respectively. The color indicates their components projected to the Sn-As spin states in the alternating spin directions along those of the adjacent Eu layers. 
The spin splitting of $\sim$0.1 eV is obtained (see also Fig. S4 b), indicative of a strong coupling between the magnetization of the Eu layer and the superconductivity of the Sn-As layer.  
{\bf d, e,} Schematic images for the mechanism of superconducting transport along the $a-b$ plane with a two-step transition: with decreasing temperature, superconducting puddles are locally formed first below $T_c$ ({\bf d}), then the phase coherence in bulk is eventually established over the whole in-plane around $T_0$ ({\bf e}). 
{\bf f,} Schematic image for the electrical transport in the $c$-axis below $T_c$; 
a finite resistance is remained due to the intrinsic magnetic Josephson junctions, which work as different phase shifters ether of 0 or $\pi$ across the ferromagnetic Eu layers with inhomogeneous domains. 
{\bf g,} An equivalent diagram of DC superconducting quantum interference device (SQUID) with the 0- and $\pi$-junctions, which cancel the phase coherence, preventing the flowing of the superconducting current.
}
\end{figure}

\clearpage

{\bf METHODS}

 {\bf  Samples}\\
Single crystals of EuSn$_2$As$_2$ were synthesized as described elsewhere by Sakagami $et~al.$ \cite{Sakagami}.
The lattice parameters determined by the powder X-ray diffraction were $a$ = 4.1993(17) \AA, and $c$ = 26.418(8) \AA~ at room temperature.

 {\bf  X-ray magnetic scattering measurement}\\
Resonant soft X-ray scattering measurements were performed at BL-16A, Photon Factory, KEK, Japan, by utilizing the horizontally polarized X-ray in resonance with the $M_5$ absorption edge of Eu (1.125 keV) \cite{NakaoJPSC}.

 {\bf  Polarizing microscopy measurement}\\
The polarizing microscopy images were observed through a commercial polarizing microscope (BXFM, Olympus) combined with an objective lens with a working distance of 33 mm (M Plan Apo, Mitutoyo). 
To obtain bright images, we used a 100 W halogen lamp (U-LH100L-3, Olympus). 
The sample was placed in a vacuum chamber and cooled using a two-stage cryocooler. 
The sample temperature was controlled with an accuracy of 0.1 K by thermal conduction from the stage of the cryocooler and a heater mounted on the sample stage \cite{TokunagaRSI,TokunagaRSI2}. 

 {\bf  Magnetization and transport measurement}\\
Magnetization and resistivity were measured down to 2.0 K using Magnetic Property Measurement System (Quantum Design) and Physical Properties Measurement System (Quantum Design), respectively. For the transport measurements at low temperatures (down to 0.1 K), we used Triton400 dilution refrigerator (Oxford). The transport properties were measured by a conventional four-terminal method with electrodes formed by room temperature curing silver paste. 

 {\bf  ARPES set-up}\\
Synchrotron-based ARPES measurements were performed at the high-resolution ARPES branch of beamline I05 of the Diamond Light Source, equipped with a Scienta R4000 analyser. 
The angular resolution was 0.2° and the overall energy resolution was better than 20 meV. 
Laser-based ARPES measurements were performed at the Institute for Solid State Physics, The University of Tokyo. The ARPES system \cite{OtaPRL} is equipped with a laser of 6.994 eV,  sixth harmonics of Nd:YVO$_4$ quasi-continuous-wave laser, and with a Scienta HR8000 electron analyzer (VG-Scienta). 
The angular and energy resolutions of measurements were set to 0.1$^\circ$ and 1.2 meV.  

 {\bf  Ab-initio calculations}\\
All of the density-functional theory (DFT) calculations were performed by VASP (Vienna Ab initio Simulation Package) \cite{VASP} using the projector augmented wave method \cite{PAW} for the pseudopotential and Pedrew-Burke-Emzerhof functional \cite{GGAPBE} for the exchange-correlation effect.
The spin-orbit coupring was included and the on-site Coulomb interaction among the Eu $f$ electrons were considered with the DFT+U method \cite{DFTU} as implemented in VASP. The on-site Coulomb interaction U was determined to be 4.0 eV for AFM state 
so that the energy level of the $f$-electron states agree with that observed in the photoemission experiment. 
To treat the magnetically ordered states, we took a doubled triclinic unit cell containing 10 atoms as discribed in Fig. S7, while we used the fundamental unit cell with 5 atoms for the non-magnetic calculations. 
On comparing the relative stability of the magnetic orientations, we fully relaxed the ionic positions starting from the experimental value \cite{Arguilla}. 
We note that to describe the paramagnetic state (spin moments at Eu sites are nonzero but disordered), we approximated it as non-magnetic (Eu spin moments are zero), as the direct treatment of randomly oriented spin configurations is difficult. Thus, we used different U values between AFM state and non-magnetic state (4.0eV for AFM; 1.6eV for paramagnetic) due to the above approximation. 
The Wannier model was constructed with the  $s$, $p$, and $f$ states of Eu,  the $s$ and $p$ states of As, and  the $s$ and $p$ states of Sn using wannier90 \cite{wannier90}. The projected spin state analyses in Fig. 4 and Fig. S5 was conducted with the AFM phase with the spin order in the $c$-axis.

\clearpage
 {\bf  Acknowledgements}\\
We thank K. Kimura and T. Kimura for low temperature magneto-optical microscopy measurement. 
We thank T. Kato, T. Kobayashi, and T. Sato for fruitful discussions. 
We thank the Diamond Light Source for access to beamline I05 under proposals SI20445-1, contributing to the results presented here. 
The magnetic X-ray scattering experiments were performed under the approval of Public Finance-Public Accountability Collective (PF-PAC) no. 2018S2-006 and 2017G597. 
Y.Ka. and R.S. were supported by CREST, Grant Number JPMJCR16Q6, Japan Science and Technology Agency (JST). 
The work was supported by KAKENHI (Grants No. JP18H01165, JP19H02683, JP19H00651, 19H01848 and 19K21842), and by MEXT Q-LEAP. 

 {\bf  Author Contributions}\\
S.Sak. Y.Ka., and T.K. planned the experimental project. 
R.S., K.H. and Y.Ka made single crystals of EuSn$_2$As$_2$.
R.S., N.A., K.H., S.Sak., T.Y., and Y.Ka. characterized single crystals of EuSn$_2$As$_2$, measured X-ray diffraction, and performed magnetization measurement. 
Y.H. performed the X-ray magnetic scattering measurement. 
Y.Y., H.N., and H.W. supported the X-ray magnetic scattering measurement. 
S.Sak, K.Kurod., Y.Ki., and M.T. performed magneto-optical microscopy measurement. 
S.Sak, M.S., S.A., N.A., T.M., and M.Y. performed transport measurement. 
S.Sak. conducted the ARPES experiment and analysed the data. 
K.Kurod., C.B., T.H., T.N., S.K., R.N., K.Kurok., T.K.K., C.C., K.O., S.Sh and T.K. supported the ARPES experiment. 
S.Sas., R.A., S.D., and N.T. calculated the band structure. 
S.Sas. and R.A. carried out theoretical analysis. 
S.Sak., S.Sas., R.A., M.Y., and T.K. wrote the paper. 
All authors discussed the results and commented on the manuscript.

 {\bf  Author Information}\\
Correspondence to Shunsuke Sakuragi ~(email: sakuragi@issp.u-tokyo.ac.jp), Minoru Yamashita ~(email: my@issp.u-tokyo.ac.jp) and Takeshi Kondo~(email: kondo1215@issp.u-tokyo.ac.jp).

\end{document}